\documentstyle[aps,psfig]{revtex}
\begin{document}

\preprint{ICRR Report 460-2000-5\\KOBE-TH-00-04}

\title{Experimental Bounds on Masses and Fluxes of Nontopological
Solitons} 

\author{~J.~Arafune and ~T.~Yoshida}
\address{Institute for Cosmic Ray Research, University of Tokyo\\
Kashiwanoha 5-1-5, Kashiwa 277-8582, Japan}

\author{~S.Nakamura}
\address{Faculty of Engineering , Yokohama National University\\
 Tokiwadai 79-5, Hodogaya-Ku, Yokohama 240-8501, Japan}

\author{~ K.Ogure}
\address{Department of Physics, Kobe University\\
 Rokkoudaicho 1-1, Nada-Ku, Kobe 667-8501, Japan}
\date{\today}
\maketitle
\begin{abstract}
    We have re-analyzed the results of various experiments which were
    not originally interested as searches for the Q-ball or the
    Fermi-ball.  Based on these analyses, in addition to the available
    data on Q-balls, we obtained rather stringent bounds on flux, mass
    and typical energy scale of Q-balls as well as Fermi-balls.  In
    case these nontopological solitons are the main component of the
    dark matter of the Galaxy, we found that only such solitons with
    very large quantum numbers are allowed.  We also estimate how
    sensitive future experiments will be in the search for Q-balls and
    Fermi-balls.
\end{abstract}


\section{Introduction}
\label{chap:intro}

In quantum field theory there exist ``nontopological solitons'' such
as Q-balls\cite{Col,Kus}, Fermi-balls\cite{Mac,Mor}, and
neutrino-balls\cite{Hol}, the stabilities of which are based on
conservation of global U(1) charges, rather than on topological
quantum numbers. For example, Q-balls are stabilised by the
conservation of a U(1) charge\cite{Col,Kus} of scalar fields, while
Fermi-balls and neutrino-balls are stabilised by the conservation of
the number of such fermions that have Yukawa couplings with scalar
fields\cite{Mac,Mor,Hol}.  If such nontopological solitons exist, they
may solve or at least be closely related to important problems in
cosmology such as dark matter\cite{Mac,Mor}, the baryon number
asymmetry of the universe\cite{Kus,Enq,Hal,Kasuya}, and gamma-ray
bursts\cite{Hol,Kus3}.

Although the ideas of such nontopological solitons are very
attractive, the qualitative properties of them, mass scale, charge
size, typical energy scale, and cosmic abundance, are so ambiguous
that there are many orders of magnitude in the parameter space that
must be considered. It is desirable to make the parameter regions of
nontopological solitons as narrow as possible through the use of
currently available observational data. This will clarify the regions
that are to be searched through experimentation in the near future.
Although there are reports of a number of useful observational or
phenomenological analyses to examine the allowed parameter regions of
Q-balls, there does not yet exist an available analysis.  In fact,
Kusenko et al.\cite{Kuz} discussed the experimental signatures of
Q-balls and pointed out powerful detection methods for neutral
Q-balls. The {\it Gyrlyanda} experiments at Lake Baikal\cite{Bai}
reported the flux limit of neutral Q-balls applying these methods to
their monopole search experiments. The report also gave rough
estimates of bounds to be obtained with other monopole search
experiments, "Baksan" scintillators\cite{Bak} and old aged
mica\cite{Mic}.  Bakari et al.\cite{Gia} calculated the energy losses
of Q-balls in matter and concluded the various {\it MACRO} detectors
can search for charged Q-balls
\footnote{ The explicit values of the flux upper limits on Q-balls are 
not available yet.
}.
When it comes to Fermi-balls, no experimental limits on the flux have
been reported to date.

In this paper, we comprehensively discuss the flux limits of Q-balls
and Fermi-balls.  As their mass could be very large, we use various
kinds of results from direct searches for supermassive magnetic
monopoles\cite{Groom}, nuclearites, and/or heavy primary cosmic
rays\cite{Sky}. We also estimate the sensitivity of present and future
experiments for the purpose of searching for nontopological solitons.
While these experiments were mainly designed for different purposes,
we stress their importance in the specific analysis of nontopological
solitons. In the following discussion, we analyze the flux bounds
taking the charge $Z_Q$ for the case of $Z_Q=0,1,2,3,10, {\rm
and}~137$, as typical values. If the charge is larger than 137, the
Q-ball will have the electromagnetic properties at low energy similar
to the case of $Z_Q = 137$, having a geometrical cross section $\pi
R^2_Q$ with the effective radius of at least $R_Q \sim 1 \mbox{\AA}$
\cite{Gla}. For the case of Fermi-balls, we assume that the electric
charge to be large enough in order to assure its stability against
deformation and fragmentation\cite{Mor}, and that the electromagnetic
properties at low energy is the same as the case of $Z_F=137$ unless
the radius exceeds $1 {\rm \AA}$.


\section{Experimental Bounds on Q-ball}
\label{Q-ball}


\subsection{Q-ball Properties}
\label{Q-property}

In the following our arguments will be restricted to the case of thick
wall Q-balls, since most attractive Q-balls in SUSY are of this
case\cite{Kus}. We briefly review the properties of Q-balls in order
to clarify our notations and assumptions in estimating the flux
bounds.

Q-balls consist of a complex scalar field with a conserved global U(1)
charge. The scalar field, $\varphi$ vanishes outside the Q-ball,
while it forms a coherent state inside with a time-dependent
phase\cite{Col},
\begin{equation}
\varphi(t,\vec{x}) = \varphi(\vec{x}) e^{-i\omega t} .
\label{phi-x-t}
\end{equation}
In this circumstance, the Q-ball is at the lowest energy state with a
fixed global U(1) charge.  This is derived by minimizing the Q-ball
mass $M_Q$,
\begin{equation}
    M_Q = \int dx \left[ \left| \partial_t \varphi \right|^2 + \left|
    {\vec \nabla} \varphi \right|^2 + V ( \left| \varphi \right|^2 )
\right] ,
\label{M_Q-def}
\end{equation}
under the constraint on the charge $Q$,
\begin{equation}
    Q =i \int dx  \left[ \varphi^{*} \partial_t \varphi - \varphi
    \partial_t \varphi^{*} \right] .
\label{Q-def}
\end{equation}
Here we assume
\footnote{ We stress that our result does not lose generality by
making this assumption. The experimental flux limits, which we obtain
later as a function of mass of the Q-ball, are independent of this
assumption in case where Q-balls are charged. This is due to the fact
that it solely depends on the mass of the Q-ball. However, in the case
of neutral Q-balls, they are dependent on this assumption, since they
rely only on their cross section rather than mass.  Of course, the
flux limitations of neutral Q-balls were independent of the assumption
if we expressed them as a function of the cross section.  Readers, who
are interested in neutral Q-balls with another type of potential, can
easily accommodate our results by matching the mass of the Q-balls
with such of ours, that has the same cross section as them.  }
that for large $\varphi$ the scalar potential $V(\left| \varphi
\right|^2)$ is almost ``flat'', i.e., $V(\left| \varphi \right|^2)
\simeq M_S^4$, where $M_S$ is a constant with a mass dimension. (The
potential of this type is known to be present in supersymmetric
theories, in which $M_S$ is a SUSY breaking scale). In this
case\cite{Kuz}, the Q-ball mass is obtained as
\begin{equation}
M_Q = \frac{4\pi \sqrt{2}}{3} M_S Q^{3/4},  \label{M_Q}
\end{equation}
with its radius,
\begin{equation}
    R_Q = \frac{1}{\sqrt{2}} M_S^{-1} Q^{1/4} . \label{R_Q}
\end{equation}
The proportionality, $M_Q \propto Q^{3/4}$, in Eq.(\ref{M_Q}) leads to
the stability of the Q-ball for $Q$ large enough.  If we consider
a baryonic Q-ball, i.e., a B-ball,
\footnote{A B-ball is the Q-ball the conserved U(1) charge of which is baryon  
number.
}
for example, the condition of stability
against nucleon emission is $ m_{\rm proton} >
\frac{\partial}{\partial Q} M_Q$, i.e.,
\begin{equation}
    Q > 5.0 \times 10^{14} \left( \frac{M_S}{\mbox{TeV}} \right)^4 .
\label{B-stability}
\end{equation}

It is pointed out that such large Q-balls may have been created in the
early universe\cite{Kasuya,Enq,Kus2}. If this is the case and such
large stable Q-balls have survived until present day, they would
contribute toward the dark matter in the Galaxy. Their flux should then
satisfy
\begin{equation}
    F \leq F_{DM} \sim \frac{\rho_{DM}~ v}{4 \pi M_Q} \sim 7.2 \times
    10^{5} \left ( \frac{\mbox{GeV}}{M_Q} \right) {\rm
    cm^{-2}sec^{-1}sr^{-1}} ,
\label{DM-limit}
\end{equation}
where $\rho_{DM}$ is the energy density of the dark matter in the
Galaxy, $\rho_{DM} \sim 0.3 {\rm GeV/cm^3}$, and $v$ is the Virial
velocity of the Q-ball, $v \sim 3 \times 10^7 {\rm cm/sec}$. In the
following analysis for Q-balls and Fermi-balls, we assume for
simplicity the velocity of Q-balls to be $v = 10^{-3} c$, where $c$ is
the light velocity.

In order to estimate efficiencies in the detection of Q-balls with
various detectors, Kusenko et al.\cite{Kuz} classified relic solitons
into two groups according to the properties of their interaction with
matter: Supersymmetric Electrically Neutral Solitons(SENS) and
Supersymmetric Electrically Charged Solitons(SECS). In this paper we
use the terms ``SENS'' and ``SECS'' only for Q-balls.  In case of
SENS, a process similar to proton decay may occur in the thin layer of
the Q-ball's surface, when the Q-ball collides with nuclei.  The
energy release of roughly 1 GeV per nucleon is carried away by pions
through this process. We call this the ``KKST process''.  In case of
SECS, with a positive charge, the KKST process in collision with
nuclei will be strongly suppressed by Coulomb repulsion. Only
electromagnetic processes will take place in this circumstance.  A
charged Q-ball with a small velocity is accompanied by an atomic-like
cloud of electrons and interacts with matter similarly to an atom with
a heavy mass. If the charge is very large, the effective interaction
radius is approximately $1 \mbox{\AA}$, which is similar to the case
of "nuclearites"\cite{Witten,Gla}.  Negatively charged Q-balls are of
little interest to us at present. If negatively charged Q-balls happen
to exist, the detection of them should be much easier than that of
positively charged Q-balls. This is due to both the electromagnetic
process and the KKST process occurring during collision. As a result
of this, the excluded parameter region of the positively charged
Q-ball is also excluded for the negatively charged ones. For the
purpose of this paper, we will only discuss the neutral and positively
charged Q-balls.


\subsection{Bounds on Flux and Mass of Neutral Q-ball (SENS)}
\label{SENS}

As pointed out by Kusenko et al.\cite{Kuz}, a neutral Q-ball would
produce the KKST process, when it collides with a nucleus, absorbing
it and emitting pions with the total energy of approximately 1 GeV per
nucleon. If the cross section of such a process is large and
successive events of this type are detected along a single trajectory,
it will be a signal detecting a Q-ball. We note that such a Q-ball
process is different from the Rubakov effect\cite{Rub} even though
they are similar to each other. The cross section for the former
process depends little on whether the target nucleus possesses a
magnetic moment or not, while the cross section for the latter process
strongly depends on it.
\footnote{ If the nucleus has a finite magnetic moment, the cross
section of the Rubakov process should have an enhancement
factor\cite{Araf} of $\sim 1/ \beta^2 \sim 10^6$ with $\beta $ being
the relative velocity (in unit of the light velocity) between the
monopole and the nucleus.}
Since the interaction cross section of the KKST process is expected of
a geometrical size, the energy loss of SENS is determined by
\begin{equation}
\frac{dE}{dx} = -  \pi R^2_Q~v^2\rho ,
\label{eloss-sens}
\end{equation}
similar to the case of nuclearite collision with a
nucleus\cite{Witten,Gla}.  Here $R_Q$ and $\rho$ are the radius of the
Q-ball and the density of the target matter, respectively. In order
for the Q-balls to reach the detector site and penetrate the detector,
the following condition should be satisfied as:
\begin{equation}
    M_Q~\gtrsim~ {1 \over 7.3} ~\pi R^2_Q \int{\rho dx}
\label{eloss-sens2}
\end{equation}
with integration over the trajectory of Q-balls. This leads to
\begin{equation}
    M_Q~ \gtrsim~ 4.2 \times 10^{-39} \left( \frac{\rho
    L}{{\rm gr~cm^{-2}}} \right)^3 \left( \frac{{\rm TeV}}{M_S}
\right)^8 {\rm GeV},
\end{equation}
where $L$ is the path length for the Q-ball to traverse matter (air,
water, and/or rock) and to penetrate the detector.  Since we are
interested in the parameter regions of $M_S > 100 {\rm GeV}$, $\rho
\leq 10~ {\rm gr~cm^{-3}}$, $L \leq 2 R_{\rm Earth} \sim 1.3 \times
10^{10}~ {\rm cm~gr~cm^{-2}}$ and $M_Q \gg M_S$, this condition is
always satisfied in any experiments on the earth.

Most experimental searches for monopole-catalyzed proton decay (the
Rubakov effect\cite{Rub}) are also sensitive to the KKST process, and
are able to give stringent bounds on SENS flux.  Let us review the
available data, not only the already reported results of the {\it
Gyrlyanda} experiments at Lake Baikal on SENS flux\cite{Bai} but also
other typical experiments reinterpreting them to get SENS flux bounds.

The {\it Gyrlyanda} experiments reported that the flux of SENS has the bound
\cite{Bai}:
\begin{eqnarray}
F < 3.9 \times 10^{-16} {\rm cm^{-2}sec^{-1}sr^{-1}} ,
\end{eqnarray} 
if the cross section for the KKST process is $\sigma > 1.9 \times
10^{-22} \mbox{cm}^2$. This corresponds to the lower limit of the SENS
mass,
\begin{eqnarray}
M_Q > 1.0 \times 10^{21} \left( \frac{M_S}{\mbox{TeV}} \right)^4
\mbox{GeV} ,
\end{eqnarray} 
since Eq.(\ref{M_Q}) and Eq.(\ref{R_Q}) relate $\sigma$ to $M_Q$ as
\begin{equation}
    \sigma = 1.9 \times 10^{-36} \left( \frac{\mbox{TeV}}{M_S}
\right)^{8/3} \left( \frac{M_Q}{\mbox{GeV}} \right)^{2/3} \mbox{cm}^2
.
\label{sigma-m}
\end{equation}

The authors of Ref.\cite{Bai} also estimated the SENS flux limit which
will be given in the {\it BAKSAN} experiments\cite{Bak}:
\begin{eqnarray}
F < 3.0 \times 10^{-16} {\rm cm^{-2}sec^{-1}sr^{-1}} ,
\end{eqnarray}
if the cross section is $\sigma > 5.0 \times 10^{-26}~ \mbox{cm}^2$,
which corresponds to
\begin{eqnarray}
M_Q > 4.2 \times 10^{15} \left( \frac{M_S}{\mbox{TeV}} \right)^4
\mbox{GeV} .
\end{eqnarray}
Although the above two experiments give considerably stringent bounds
to SENS, we must examine how stringent limits will be given by other
monopole searches and future cosmic ray experiments.

The {\it Kamiokande} Cherenkov detector\cite{Kam}, which has 3 000
tons of water, approximately 1 000 photo-tubes and is located at
approximately 1 000 meters underground, gave the limits of monopole
with 335 days of live time.  We reinterpret the results to obtain SENS
flux limits approximately\cite{Apology}:
\begin{eqnarray}
F < 3 \times 10^{-12},~~ 3 \times 10^{-14},~ {\rm and} ~~3 \times
10^{-15}~{\rm cm^{-2}sec^{-1}sr^{-1}},
\end{eqnarray}
for SENS with $\sigma = 0.1,~~1, ~{\rm and} ~~10 ~\mbox{mb}$, which
correspond to
\begin{eqnarray}
M_Q = 4.0 \times 10^{11},~~1.2 \times 10^{13},~~{\rm and}~ 5.6 \times 10^{13}
\left( \frac{M_S}{\mbox{TeV}} \right)^4 \mbox{GeV} ,
\end{eqnarray}
respectively.

The {\it Super-Kamiokande} experiments with 50 000 tons of water, the
Cherenkov detector would obtain more stringent flux limits than those
of {\it Kamiokande} by almost two orders of magnitude and with three
years of observation time.

The {\it MACRO}, large underground detectors\cite{Gia,Macr,GiaN}
consisting of three kinds of subdetectors ( i.e. scintillation
counters, streamer tubes, and nuclear track detectors(CR-39)) could
search for SENS\cite{Gia}. Although the flux limits of SENS are
reported to be obtained as $F \lesssim~ 10^{-16} {\rm
cm^{-2}sec^{-1}sr^{-1}}$ in Ref.\cite{Gia}, we do not give this flux
limit here since the detection efficiency is too difficult for us to
estimate.

The {\it AMANDA} Cherenkov detectors\cite{Ama,Amandmono} located at
the South Pole under ice, are designed mainly to detect relativistic
particles. If they could also detect slow particles, the large area of
the array would be of great help.  For a few years to observe SENS,
they could obtain their flux limits as $F \lesssim~ 10^{-17} {\rm
cm^{-2}sec^{-1}sr^{-1}}/\epsilon$ with the detection efficiency,
$\epsilon$. The flux limits, however, are not estimated here since the
detection efficiencies are also difficult for us to evaluate.

The {\it TA}({\it Telescope Array Project}) with effective aperture of
$ S\Omega \epsilon = 6 \times 10^3 ~{\rm km^2~sr}$ \\($S\Omega = 6
\times 10^4 ~{\rm km^2~sr}$ and the duty factor $\epsilon = 0.1$) is
planned to detect cosmic rays with energy, $10^{16} - 10^{20}~
\mbox{eV}$, by detecting the air fluorescence\cite{Ta}. Using a
special trigger, it may detect slow particles with large energy
loss\cite{Sasaki}. With such a trigger, it may be possible to search
for SENS flux at the level of $F< 1 \times 10^{-21} {\rm
cm^{-2}sec^{-1}sr^{-1}}$, provided that the fluorescence light yield
of the event is equivalent to that of air showers with the minimum
energy $E_{min} = 10^{16} \mbox{eV}$. Considering energy release of
roughly 1 GeV per absorbed nucleon in the KKST process, one obtains
the condition of detectability in {\it TA}:
\begin{equation}
    \sigma \int{\rho dx} \geq E_{min}/ \xi_{SENS} ,
\label{ta-detect}
\end{equation}
where $\xi_{SENS}$ is the ratio of the efficiency of fluorescence
light yield per total energy loss of SENS to that of extensive air
showers. In the KKST process we estimate $\xi_{SENS} \sim 1,$ since
emitted pions are relativistic.  Assuming the average density of air
$\rho \sim 0.5 \times 10^{-3}~ {\rm g/cm^3}$ and the path length of
SENS in the atmosphere $L \sim 20 ~{\mbox{km}}$, we obtain the lower
limit of SENS mass from Eq.(\ref{ta-detect}),
\begin{eqnarray}
M_Q > 0.7 \times 10^{24} \left( \frac{M_S}{\mbox{TeV}}
\right)^4 \mbox{GeV} .
\end{eqnarray}

The {\it OA}({\it OWL-AIRWATCH}) detector\cite{Owl} located on a
satellite with effective aperture $ 5 \times 10^{5}~{\rm km^2~sr}$
(with duty cycle of $ 0.2 $ taken into account), is used to observe
the atmospheric fluorescence of air showers with energy $10^{19} -
10^{20}$ eV or more.  If it could also detect slow particles with high
efficiency, it would be possible to search for SENS during a one year
time constraint at the flux level of $F < 10^{-23} {\rm
cm^{-2}sec^{-1}sr^{-1}}$. For SENS to yield the amount of fluorescence
light comparable to that of air showers with $10^{19}$eV, their mass
should satisfy $M_Q > 2 \times 10^{28} \left ( \frac{M_S}{\mbox{TeV}}
\right)^4 \mbox{GeV}$.

The bounds given in the various experiments mentioned above are
summarized in Fig.\ref{z0com}(a). This figure also gives the flux to
be expected if the dark matter of the Galaxy consists mainly of
SENS(see Eq.(\ref{DM-limit})).  Note that these bounds of excluded
regions depend on $M_S$. ( The lower mass bound in each experiment is
calculated taking $M_S = 1 \mbox{TeV}$ in this figure. In case of {\it
TA} and {\it OA}, we added those for $M_S=100~ {\rm GeV}$. ) The
region in which it is expected that future experiments are able to be
conducted, are also given there.  If we assume that the dark matter of
the Galaxy consists mainly of SENS ( i.e. if the flux is just on the
DM-limit line of Fig.\ref{z0com}(a) ), we obtain the excluded regions
of SENS mass as functions of the parameter $M_S$ as in
Fig.\ref{z0com}(b).  This figure shows that SENS with the U(1) charge
$Q~ \gtrsim~ 10^{22}$ are cosmologically interesting in the region of
$M_S~ \gtrsim~ 6 \times 10^{3} {\rm GeV},$ while that {\it TA} and
{\it OA} could search for SENS of $10^{25}~
\lesssim~Q~\lesssim~10^{35}$ in the region of $M_S~ \lesssim~ 6 \times
10^{3} {\rm GeV}.$

\subsection{Bounds on Flux and Mass of Charged Q-ball (SECS)}
\label{SECS}

Charged Q-balls (SECS) interact with matter in a similar way to
nuclearites\cite{Witten,Gla}. The rate of energy loss of SECS in
matter is
\begin{equation}
\frac{dE}{dx} = - \sigma \rho v^2 ,
\label{energy-loss}
\end{equation}
where $\sigma$ is the cross section of SECS collision with matter,
$\rho$ is the density of the matter, and $v$ is the relative velocity
of SECS and the matter. Substituting $E = \frac{1}{2}M_Q v^2$ in the
above equation and integrating with respect to $x$, one obtains
\begin{equation}
\rho L = \frac{M_Q}{\sigma} \ln{\frac{v_0}{v_c}} ,
\label{rho-L}
\end{equation}
where $L$ is the range of SECS in the medium, $v_0$ is an initial
velocity of SECS($\sim 3 \times 10^7 {\rm cm/sec}$), and $v_c$ is a
final velocity of SECS. We use the estimation in the case of
nuclearites, $v_c \sim 1.7 \times 10^4 ~{\rm cm/sec}$ in
rock\cite{Gla}.  SECS with a velocity below this value are quickly
brought to rest. In the case of SECS with a radius larger than $1
\mbox{\AA}$ ( i.e., $M_Q > 2.1 \times 10^{30} \left( M_S/\mbox{TeV}
\right)^4 {\rm GeV}$ ), the cross section is simply given by $\sigma =
\pi R_Q^2$.  From Eqs.(\ref{M_Q}), (\ref{R_Q}) and (\ref{rho-L}), we
obtain
\begin{equation}
    \rho L = 6.2 \times 10^{12} \left( \frac{M_S}{\mbox{TeV}}
\right)^{8/3} \left( \frac{M_Q}{\mbox{GeV}} \right)^{1/3} {\rm gr/
cm^2} .
\label{rho-L1}
\end{equation}
Such heavy SECS should, however, have too small a flux to be detected
according to Eq.(\ref{DM-limit}), and thus, are not interesting to us.
We have considered only SECS with a radius smaller than $1 \mbox{\AA}$
in the following.

If such SECS have a significantly large electric charge ($Z_Q \gtrsim~
\alpha^{-1} \sim 137$ where $\alpha$ is the fine structure constant),
the cross section of their collision with matter is not controlled by
their intrinsic radius $R_Q$, but by the size of the surrounding
electron cloud. This cloud is never smaller than $\sim 1
\mbox{\AA}$. We note that this situation is similar to the case of
nuclearites\cite{Gla}.  We have $\sigma = \pi R_{eff}^2 $ with
$R_{eff} =1 {\mbox{\AA}}$, which leads to the relation
\begin{equation}
    \rho L = 4.0 \times 10^{-8} \left( \frac{M_Q}{\mbox{GeV}} \right)
    {\rm gr/ cm^2} .
\label{rho-L2}
\end{equation}
Equation (\ref{rho-L2}) shows that SECS with $Z_Q ~\gtrsim~ 137$
should be as heavy as
\begin{equation}
    M_Q > 2.5 \times 10^{7} \left( \frac{\rho L}{{\rm gr / cm^2}}
\right) \mbox{GeV}
\label{rho-L3}
\end{equation}
in order to penetrate the medium with density $\rho$ and length $L$.

If the charge of SECS is small, i.e. $Z_Q \ll 137$ (and the intrinsic
SECS radius $R_Q$ is also smaller than $1\mbox{\AA}$) , the effective
cross section should be smaller than $ \pi {\rm (1 \AA)^2}$. We then
need more delicate treatments to estimate the rate of the energy loss.
It is known that there are two kinds of interaction that contribute to
the energy loss of SECS at low velocity $\beta \sim 10^{-3}$ (
i.e. interaction with electrons and interaction with nuclei ), $dE/dx
= (dE/dx)_{\rm electrons} + (dE/dx)_{\rm
nuclei}$\cite{Gia,Lind,Phip,Cano,Wil,Fice}.  The rate of electronic
energy loss is estimated as Ref.\cite{Lind},
\begin{equation}
    \left( \frac{dE}{dx} \right)_{\rm electrons} = 8 \pi ~\alpha a_0
        \frac{v}{v_0} N_e \frac{Z_Q^{7/6}}{ \left( Z_Q^{2/3} + Z^{2/3}
    \right)^{3/2}} ,
\label{loss-e}
\end{equation}
where $v_0$ is given by $\alpha c$, $N_e$ is the number density of
electrons in the medium, $Z$ is the atomic number of the medium, and
$a_0$ is the Bohr radius.  The rate of energy loss caused by
interaction with the nuclei of the medium is given in Ref.\cite{Wil}
\begin{equation}
    \left( \frac{dE}{dx} \right)_{\rm nuclei} = 4 \pi \alpha a N_Z Z_Q
        Z \frac{M_Q}{M_Q+M} \frac{A~ \ln{B \epsilon}}{B \epsilon - (B
        \epsilon)^{-C}} ,
\label{loss-n}
\end{equation}
with $A=0.56258$, $B=1.1776$, $C=0.62680$, $M$ is the mass of the
target nucleus, ~$N_Z$ is the number density of target nuclei and
\begin{equation}
    \epsilon = \frac{a M~M_Q \beta^2}{2 \alpha Z_Q Z(M_Q+M)} , ~~~a =
    \frac{0.8853 a_0}{\left( \sqrt{Z_Q} + \sqrt{Z} \right)^{2/3}} .
\end{equation}
In case of $M_Q \gg M$ we calculated $(dE/dx)_\rho ~ \equiv~
(dE/dx)/\rho$ (which depends only loosely on the density of the
medium) for ${\rm SiO_2}$ and obtained $(dE/dx)_\rho = 0.16,~ 0.41, ~
0.71$, and $2.9 {\rm GeV/gr/cm^2}$ for SECS with typical charge $Z_Q =
1, ~2, ~3, ~{\rm and} ~10, $ respectively.  We also find the energy
loss rate $(dE/dx)_{\rho}$ depends little on matter medium(air, water,
and/or rock). We use the above values for the following estimation of
energy loss.  The condition for SECS to penetrate the medium with
density $\rho$ and length $L$ is roughly given by
\begin{equation}
    \frac{1}{2}M_Q \beta^2 > \left( \frac{dE}{dx} \right)_{\rho} \rho
    L .
\label{rho-L4}
\end{equation}

Since SECS have a large cross section when they collide with matter,
they should have kinetic energy large enough to reach a detector and
to penetrate it.  This situation is unlike that of SENS.  Let us take
the following experiments as typical in detecting SECS.

The {\it MACRO} with its three subdetectors are sensitive to
SECS\cite{Gia,Macr,GiaN} for any values of $Z_Q$; in order for SECS to
reach the underground detector site from above with $\rho L \sim 3.7
\times 10^5 {\rm gr~cm^{-2}}$, the lower limit of the SECS mass $M_Q$
is given by Eq.(\ref{rho-L3}) and Eq.(\ref{rho-L4}), as $M_Q >1.2
\times 10^{11},~3.0 \times 10^{11},~5.3 \times 10^{11},~2.1 \times
10^{12}, ~{\rm and} ~9.3 \times 10^{12}$ GeV for $Z_Q = 1, 2, 3, 10, ~
{\rm and} ~ 137,$ respectively.  The flux limit suggested by the
results of no events of monopole search experiments\cite{Gia} is
roughly\cite{Apology} $F~\lesssim~ 10^{-16}{\rm
cm^{-2}sec^{-1}sr^{-1}}$.  It is obvious that the inclusion of
searches for Q-balls from below, decreases the flux upper limit by a
factor of two, although the lower limits for the mass to reach the
detector increase by a factor of $(\rho L)_{{\rm from ~below}}/ (\rho
L)_{{\rm from~above}} = (6.6 \times 10^9) / (3.7 \times 10^5 ) \sim
1.8 \times 10^4$ .  This feature for SECS is common to other
experiments such as {\it OYA}, {\it NORIKURA}, {\it KITAMI}, {\it
AKENO}, {\it UCSD}$I\!I$, {\it KEK}, and {\it MICA}.

The {\it OYA} experiments\cite{Oya} with CR-39 plastic track detectors
(located at the depth of $\rho L = 10^4{\rm cm^{-2}sec^{-1}sr^{-1}}$),
searched for monopoles and nuclearites for 2.1 years. The flux limit
of these experiments corresponds to that of SECS, $F < 3.2 \times
10^{-16}{\rm cm^{-2}sec^{-1}sr^{-1}}$\cite{Apology}. Since the
detectors are sensitive to the restricted energy loss larger than $
\sim 0.14 {\rm GeV~gr/cm^2}$, it is possible that SECS with a charge
of $Z_Q \geq 2 $ may be detected.  The condition for SECS to reach the
detector is $M_Q >8.2 \times 10^{9}~{\rm GeV}, ~1.4 \times
10^{10}~{\rm GeV}, ~5.8 \times 10^{10}~{\rm GeV}, ~{\rm and} ~ 2.5
\times 10^{11}~{\rm GeV}$ for $Z_Q = 2, ~3, ~10, {\rm and} ~137,$
respectively.

The {\it KEK} experiments with scintillation counters placed at ground
level\cite{Kek} should be sensitive to SECS at any value of $Z_Q$. As
the detection threshold is $0.01~I_{min}$ where $I_{min}$ is the
minimum energy loss of the ionizing particle with $Z=1$\cite{Apology}.
We interpret that their flux upper limit for strange quark matter
(nuclearites) holds for the SECS: $F < 3.2 \times 10^{-11}{\rm
cm^{-2}sec^{-1}sr^{-1}}$. The mass bounds for SECS to reach detectors
located at ground level are given by $M_Q > ~3.1 \times 10^8 ~{\rm
GeV}, ~8.2 \times 10^8 ~{\rm GeV}, ~1.4 \times 10^9 ~{\rm GeV}, ~5.8
\times 10^9 ~{\rm GeV}, ~{\rm and} ~2.5 \times 10^{10} ~{\rm GeV}$ for
$Z_Q =1,~2,~3,~10~, {\rm and}~137$, respectively.

The {\it KITAMI} experiments with CN (cellulose nitrate) nuclear track
detectors installed in houses at sea level, give the flux limit $F <
5.2 \times 10^{-15} {\rm cm^{-2}sec^{-1}sr^{-1}}$\cite{Kit}. We
reinterpret this as the SECS flux limit\cite{Apology}.  Since the
ionization energy loss of SECS should be larger than $\sim 1.3 {\rm
GeV/ gr /cm^{2}}$ for a chemically etchable track to be made in CN, we
estimate that SECS with $Z_Q \geq 10$ can be detected with the CN
detectors.  At ground level with $\rho L = 10^3 ~{\rm gr~cm^{-2}}$,
SECS heavier than $ 5.8 \times 10^{9}~ {\rm GeV}~ {\rm and} ~2.5
\times 10^{10}~ {\rm GeV} $ can penetrate the atmosphere and reach the
detector from above. This is the case for $Z_Q =10~ {\rm and} ~137$,
respectively.

The {\it AKENO} experiments with Helium-gas counters\cite{Ake} located
at ground level should be sensitive to SECS at any value of $Z_Q$. The
detector is composed of many layers of proportional counters and
concrete shields. The detector is sensitive to tracks with an
ionization level larger than $10 ~I_{\rm mim}$. This should therefore
assure the detectability for SECS with any value of $Z_Q$.  SECS with
masses of $M_Q > 5.8 \times 10^{8} {\rm GeV},~1.5 \times 10^{9} {\rm
GeV},$ $~2.6 \times 10^{9} {\rm GeV}$ $,~1.1 \times 10^{10} {\rm
GeV},~{\rm and}~4.7 \times 10^{10}{\rm GeV}$, can penetrate the earth
and the concrete shields of the detector from above for $Z_Q =
1,~2,~3,~10,~{\rm and}~ 137$, respectively. Due to the fact that no
events have been detected, we estimate the flux limit as $F < 1.8
\times 10^{-14} {\rm cm^{-2}sec^{-1}sr^{-1}}$\cite{Apology}.

The {\it UCSD}$I\!I$ experiments with He-CH${}_4$ proportional tubes
placed at ground level\cite{Ucs}, should be sensitive to SECS with any
value of $Z_Q$, as the detection threshold is $0.09 ~I_ { mim}$.  The
flux upper limit for monopoles $F = 1.8 \times 10^{-14} {\rm
cm^{-2}sec^{-1}sr^{-1}}$ should be the same as that for
SECS\cite{Apology}.  The mass bounds are the same as in the above case
of the {\it KEK} experiments.

The {\it NORIKURA} CR-39 experiments were conducted at the top of
Mt.Norikura in the search for monopoles and strange quark
matter\cite{Nor}.  They are sensitive to tracks with ionization larger
than $0.35 {\rm GeV/gr/cm^2}$. This assures that SECS with $Z_Q \geq
3$ will be detectable.  Since the detector is installed at a high
altitude, it is more sensitive to lighter SECS compared to detectors
such as those used by {\it KITAMI}, {\it AKENO}, {\it UCSD}$I\!  I$,
and {\it KEK}, which are installed at ground level. The flux limit is
$F = 2.2 \times 10^{-14} {\rm cm^{-2}sec^{-1}sr^{-1}}$.

The {\it MICA} analysis with ancient mica crystals of $0.6 - 0.9
\times 10^9$ years old, was made to search for monopoles\cite{Mic}.
It should also be sensitive to SECS with $Z_Q \gtrsim~ 10$ as the
detection threshold is $2.4 {\rm GeV/gr/cm^2}$ \cite{Apology}. In
order for SECS coming from above to reach the mica crystals located at
3 km underground with $\rho L = 7.5 \times 10^{5} {\rm gr/ cm^2}$,
SECS should be heavier than $4.4 \times 10^{12} {\rm GeV} ~{\rm
and}~1.9 \times 10^{13} {\rm GeV} $ for $Z_Q =10~{\rm and}~ 137$,
respectively. In case of the monopole search, the capture of an
aluminium atom by a monopole was taken into account. The detection
efficiency was estimated at $\sim 0.15$~for monopoles coming from
above and 0 for those from below, due to this consideration.
Therefore, this decrease of efficiency caused the flux limits to be
less stringent.  In case of SECS search, however, SECS need not
capture an aluminum atom and can be heavy enough to penetrate the
earth. With these considerations we estimate the flux limit for SECS,
which is better than that for monopoles by a factor of 6, to have $F =
2.3 \times 10^{-20} {\rm cm^{-2} sec^{-1} sr^{-1}}$ for SECS from
above and a value improved by a factor of twelve for SECS from any
direction.

The {\it SKYLAB} experiments with Lexan track detectors (installed in
the SKYLAB workshop located in space), reported that no superheavy
relativistic nuclei ($Z \gtrsim~ 110$)\cite{Sky} had been detected,
which correspond to $Z_Q ~\gtrsim~ 10$ in our case of slow particles.
The Lexan detector and its container have a thickness $\rho L = 2 {\rm
gr~cm^{-2}}$ ~in total, or effectively $\rho L \sim 3 {\rm
gr~cm^{-2}}$, giving the mass a lower limit of SECS $ M_Q > 1.7 \times
10^{7} {\rm GeV}~ {\rm and} ~7.5 \times 10^{7} {\rm GeV} $ for $Z_Q
=10 ~ {\rm and} ~ 137$, respectively. This experiment gives the flux
limit of SECS as $F < 3.8 \times 10^{-12} {\rm
cm^{-2}sec^{-1}sr^{-1}}$\cite{Apology}.

The {\it AMS} experiments may detect SECS in the future if a special
trigger designed to detect slow particles is used\cite{Ams}. The
magnetic spectrometer on the space station would have a large area,
allowing us to record the flux upper limit as $F \times 10^{-11} {\rm
cm^{-2}sec^{-1}sr^{-1}}$ given a one year observation time
frame\cite{Apology}.  The detection with thickness of $\rho L \sim 10
{\rm gr~cm^{-2}}$ ~needs mass $M_Q > 3.1 \times 10^{6} {\rm GeV},~8.2
\times10^{6} {\rm GeV}$, $~1.4 \times 10^{7} {\rm GeV},~5.8 \times
10^{7} {\rm GeV}, ~{\rm and} ~2.5 \times 10^{8} {\rm GeV}$ for $Z_Q
=1,~2,~3,~10 ~ {\rm and} ~ 137$, respectively.

The experimental data and also the future possibilities mentioned
above are summarized in Figures\ref{z1com}-\ref{z137com}.  Figures
\ref{z1com}(a)-\ref{z137com}(a) represent limits of the flux and the
mass for SECS.  Figures\ref{z1com}(a)-\ref{z3com}(a) show that future
{\it AMS} experiments and current {\it MACRO} experiments are very
sensitive to a wide parameter region of mass, charge and flux of SECS.
On the other hand, Figure\ref{z10com}(a) and Figure\ref{z137com}(a)
show that the {\it SKYLAB} and {\it MICA} experiments give us a
stringent exclusion of the lower and upper region of mass
respectively. Figures\ref{z1com}(b)-\ref{z137com}(b) represent bounds
on the U(1) charge $Q$ of SECS, versus the symmetry breaking parameter
$M_S$ in the case that SECS are mainly contributing to dark matter in
the Galaxy. If we assume that SECS are B-balls and impose their
condition of stability, these figures show that only SECS with
$Q~\gtrsim~10^{22-26}$ remain to be considered.

\section{Experimental Bounds on Fermi-ball}
\label{Fermi-ball}


\subsection{Fermi-ball Properties}
\label{F-property}

We have investigated which regions are allowed for the flux of
Fermi-balls and discuss our findings in this section. The Fermi-ball,
which is another kind of nontopological soliton, was first proposed by
Macpherson and Campbell\cite{Mac} as a possibility for explaining the
make-up of the dark matter in Our Galaxy.  Fermi-balls with a large
radius are found to be unstable against deformation and are therfore
expected to fragment into very small Fermi-balls\cite{Mac}.  The
Fermi-ball then interacts with matter in a weak manner and seems too
difficult to be detected.  It was then proposed by Morris\cite{Mor}
that Fermi-balls be electrically charged in order to improve its
stability against deformation and fragmentation.  Although a
Fermi-ball with large fermion number is energetically unstable and
could fragment into many smaller Fermi-balls, the Coulomb force is
expected to suppress the process of fragmentation, increasing the life
time of the large Fermi-ball so that it would become more stable.
Owing to the electromagnetic interaction, various detectors become
sensitive to Fermi-balls.  In this paper we have focused on the
experimental bounds for charged Fermi-balls rather than on neutral
Fermi-balls.

Let us first briefly review the properties of Fermi-balls in order to
make our assumptions and terminologies clear.  The charged Fermi-ball
consists of three components: a scalar field, a large number of
electrically charged heavy fermions, and also a large number of
electrons (or positrons) which partly compensate for the electric
charge of the heavy fermions.  (When we consider the stability of
Fermi-balls in the following, we assume that the electric charge of
the heavy fermions is positive without loss of generality.)  Inside
the Fermi-ball, the scaler field has the value of a false vacuum 
(which is almost degenerate to the true vacuum
\footnote{The approximate degeneracy of energy densities of a true and
false vacuum is necessary for the stability of large Fermi-balls.}
), while outside the Fermi-ball, it has the value of a true vacuum.
On the boundary wall, the energy density of the scalar field is higher
than those in the two vacua. Since the heavy fermions have a large
mass in both the true and false vacuum and have a vanishing mass on the
wall, many fermions are strongly trapped on the wall.

Since the electric field of the Fermi-ball is strong,
electron-positron pairs may be created by the effect of the quantum
field theory. This effect decreases the electric field strength by
leaving the electrons on the surface and emitting the positrons to
infinity.  Assuming the difference of the energy density between two
vacua is small enough to be neglected, we obtain the energy of the
Fermi-ball with the number of heavy fermions $N_F$ and the number of
electrons $N_e$,
\begin{eqnarray}
    E_F = 4 \pi \Sigma R^2 +
    \frac{2(N_F^{\frac{3}{2}}+N_e^{\frac{3}{2}})}{3R} + \frac{\alpha
    (N_F-N_e)^2}{2 R} .
\label{E_F}
\end{eqnarray}
Here, $\Sigma$ is the surface tension of the Fermi-ball and $R$ is the
radius of it.  The radius of the stable Fermi-ball, $R_F$ is
determined by balance 
\footnote { We have two ways in differentiating Eq.(\ref{E_F}) with
respect to $R$.  : either one fixes $N_F$ and $N_e$ , or one fixes
$N_F$ and the electric field ${\cal E}$ on the surface of the
Fermi-ball has the critical value ${\cal E} = m_e^2/e$. Here we take
the former and follow Morris\cite {Mor}. For the other possibility,
see Ref.\cite{Prepare}.  }
of the surface tension energy (the first term) proportional to $R^2$,
and the Fermi energy (the second term) or the Coulomb energy (the
third term) which is proportional to $R^{-1}$ as
\begin{eqnarray}
    \label{r0}
    R_F = \left\{ \frac{1}{8\pi \Sigma}\left
    [ \frac{2(N_F^{\frac{3}{2}}+N_e^{\frac{3}{2}})}{3} + \frac{\alpha
    (N_F-N_e)^2}{2 R}\right] \right\}^{\frac{1}{3}}.
\end{eqnarray}
The energy of the Fermi-ball is then,
\begin{eqnarray}
    E_F~ ( = M_F ) = 12\pi \Sigma R_F^2 \equiv \kappa^3 R_F^2~ ,
\label{efmin}
\end{eqnarray}
which is the common relationship when the volume energy is neglected.
Our following analyses of the Fermi-ball's flux are based only on the
above relation.  Therefore, the experimental limits from these
analyses do not depend on the details of Fermi-ball models.


\subsection{Bounds on Flux and Mass of Fermi-ball}
\label{F-ball}

Throughout this section, we discuss the physical parameter regions
that have been excluded in experiments.  If the electric charge of the
heavy fermions is positive, the observational situation is almost the
same as for the case where SECS has $Z_Q ~\gtrsim~ 137$.  This means
that the effective radius can be taken as $R_{eff}\sim 1\mbox{\AA}$
when the intrinsic radius is smaller than this value. In the case of
the Fermi ball (not as is the case with the Q-ball), however, the
radius can be larger than $1\mbox{\AA}$ without becoming too heavy,
as its mass is proportional to $R_F^2$ rather than $R_F^3$.  The
effective radius thus becomes,
\begin{eqnarray}
    R_{eff} = \left\{
      \begin{array}{rl}
          1\mbox{\AA} &\quad ({\rm  for}~R_F<1\mbox{\AA})\\
          R_F  &\quad ({\rm  for}~R_F\geq 1\mbox{\AA}) .\\
          \end{array}
    \right.
\label{r}
\end{eqnarray}
In the case where $R_F$ is large enough, Fermi-balls can be detected
with not only detectors which are sensitive to SECS with $Z_Q \gtrsim
137$ , but also with future detectors used for extensive air showers
such as {\it TA} and {\it OA}.  This detectability within future
experiments is the main difference in the case of SECS with $Z_Q
\gtrsim 137$.  We examine the bounds on Fermi-balls
\footnote {In the following, we assume that the electric charge of
Fermi-balls is positive. In case Fermi-balls have a negative electric
charge, it is much easier for us to detect them than those with a
positive charge. This is followed by the fact that the negatively
charged Fermi-balls may trap nuclei in collision with matter and emit
mesons or photons with total energy of order 1GeV per nucleon.

We note that all regions excluded for the positively charged
Fermi-balls should also be excluded for the negatively charged
Fermi-balls.  }
to be given by {\it AMS, SKYLAB, UCSD}$I\!I$, {\it AKENO, KEK,
NORIKURA, KITAMI, OYA, MACRO, MICA, TA}, and {\it OA} in the
following.

First, we discuss the conditions in which the Fermi-ball reaches and
penetrates the detectors. For Fermi-balls with $R_{eff}=1\mbox{\AA}$
($R_F < 1\AA$), the condition is the same as that for SECS with $Z_Q
\gtrsim 137$, i.e. Eq.(\ref{rho-L3}),
\begin{equation}
M_F > 2.5 \times 10^{7} \left( \frac{\rho L}{{\rm gr/ cm^2}}
\right) {\rm GeV} .
\label{M-cond}
\end{equation}
For Fermi-balls with $R_{eff}=R_F$ ($R_F > 1\AA$ ), the condition is
different from that of SECS (Eq.(\ref{rho-L1})), since the relation
between the mass and the radius of Fermi-balls is different from that
of Q-balls. Using Eq.(\ref{rho-L}) and Eq.(\ref{efmin}) with $\sigma =
\pi R_F^2$, we obtain the condition
\begin{eqnarray}
    \kappa
    \geq
    4.7\times 10^{-2}
    \left(
      \frac{\rho L}{{\rm gr/ cm^2}}
    \right)^{\frac{1}{3}} {\rm GeV} .
\label{kappa-cond}
\end{eqnarray}
This condition is independent of the mass of Fermi-balls in the case
of $R_F \geq 1 ~ \mbox{\AA}$. From Eq.(\ref{M-cond}) and
Eq.(\ref{kappa-cond}) we see that all the experiments (except {\it TA}
and {\it OA}) give the same bounds on mass and flux of Fermi-balls as
on those of SECS with $Z_Q \geq 137$, as the detection efficiencies
for these two kinds of solitons are the same for these detectors.

We next discuss the conditions for detecting Fermi-balls with {\it TA}
and {\it OA}. This requires a fluorescence light yield which
corresponds to the air shower energy of $E_{min}=10^{16} \mbox{eV}$
for {\it TA}\cite{Ta,Sasaki} and $E_{min}=10^{19 }\mbox{eV}$ for {\it
OA}\cite{Owl}.  This condition is satisfied if the energy loss of the
Fermi-ball is
\begin{equation}
    \pi R_F^2 \rho v^2 L
    \geq
    E_{min}/ \xi_F .
\label{eloss-ta-owl}
\end{equation}
Here, the parameter $\xi_F$ is the ratio of the efficiency of
fluorescence light yield per total energy loss for slow Fermi-balls to
that for high energy cosmic ray protons. We estimated $\xi_F$ from the
measurements of the efficiency of ionization for slow
ions\cite{Cano,Phip} as $\xi_F \sim 1/5$.  By taking the average
density of air as $\rho=5.0\times 10^{-4} {\rm g/cm^3}$, the velocity
of the Fermi-ball as $v=10^{-3} c$ and the length of the trajectory as
$L=20 ~{\rm km}$, one obtains the required condition to observe
Fermi-balls from Eq.(\ref{efmin}) and Eq.(\ref{eloss-ta-owl})
\footnote {Here we did not assume the black body radiation from the
Fermi-ball trajectory,  since it is effective only for dense medium\cite{Gla}. },
\begin{eqnarray}
    M_F \geq 6.5 \times 10^{22} \left( \frac{\kappa}{10^3~ {\rm GeV}}
\right)^3 \left( \frac{E_{min}}{10^{16} {\rm eV}} \right) {\rm GeV} .
\label{tacon}
\end{eqnarray}

If {\it TA} can observe slow particles
\footnote{The {\it TA} experiments may be available for slow particle
search with a special trigger\cite{Sasaki}},
it will be able to search for Fermi-balls on order to give the same
flux limit as that of SENS, $F< 1 \times 10^{-21} {\rm
cm^{-2}sec^{-1}sr^{-1}}$\cite{Ta}. Equation(\ref{tacon}) gives the
mass bounds required to observe Fermi-balls, $ M_F \geq 6.5 \times
10^{22} \left ( \frac{\kappa}{10^3~ {\rm GeV}} \right)^3 {\rm
GeV}$. If {\it OA} can observe slow particles as well, the flux bounds
to be obtained are improved by two orders of magnitude $F< 1 \times
10^{-23} {\rm cm^{-2}sec^{-1}sr^{-1}}$\cite{Owl}, for $M_F \geq 6.5
\times 10^{25} \left ( \frac{\kappa}{10^3~ {\rm GeV}} \right)^3 {\rm
GeV}$.

The flux limit with $\kappa =10^3$ GeV~ is shown in
Figure\ref{fermicom}(a).  This figure shows that quite broad regions are
already excluded by available experimental data. Future experiments,
{\it TA} and {\it OA}, may find it possible to search large regions
which have not been previously accessible through current experiments.
Figure\ref{fermicom}(b) shows the region of $M_F-\kappa$ plane
(hatched with solid lines and with half-tone dot meshing) to be
excluded when we assume that the dark matter of the Galaxy consists
mainly of Fermi-balls.

Here we discuss how these results constrain the Morris's simple
Fermi-ball model\cite{Mor}.  The electric field becomes the critical
value (${\cal E}= m_e^2/e$), near the surface due to the surrounding
electrons in this model. The third term of Eq.(\ref{E_F}) is $m_e^4
R^3/2 \alpha$ in this case. When this Coulomb energy is related to the
surface energy as $4 \pi \Sigma R^2= C m_e^4 R^3/2 \alpha$
\footnote{ Since the surface energy must be the same order as the
total energy, $C$ must not be much smaller than unity.  Investigation
into what range of $C$ is allowed to stabilize the Fermi-ball is
underway\protect\cite{Prepare}.  }
we obtain,
 \begin{eqnarray}
     M_F \sim 3(4\pi \Sigma)^3 \left(\frac{2\alpha}{C m_e^4}\right)^2
     = 6.0\times 10^{21} C^{-2}\left(\frac{\kappa}{{\rm GeV}}\right)^9
     {\rm GeV} .
\label{mmodel}
\end{eqnarray}

In order to prevent the Fermi-ball from forming a black hole, its
radius should be larger than the Schwarzschild radius, $M_F/M_{pl}^2$,
where $M_{pl}$ is the Planck mass.  From this condition, we obtain the
following constraint of $\kappa$ using Eq.(\ref{mmodel}),
\begin{eqnarray}
    \kappa \leq 3.3\times 10^4 C^{\frac{1}{6}}{\rm GeV}.
\label{bh2}
\end{eqnarray}


We draw lines which represent Eq.(\ref{mmodel}) and Eq.(\ref{bh2}) in
Figure\ref{fermicom}(b).  We find that the {\it TA} and {\it OA}
experiments are powerful to search the new region of approximately
$10^{17}{\rm GeV}~\lesssim ~M_F~\lesssim ~10^{29}{\rm GeV}$ in
Morris's Model (taking $C=1$).


\section{Conclusion}
\label{concl}

Quantum field theory allows the existence of such nontopological
solitons as Q-balls and Fermi-balls, the stability of which is
supported by the conservation of a global U(1) charge.  These solitons
may play important roles in cosmology in solving the problems of dark
matter, baryogenesis, and gamma ray bursts. In this paper, we have
considered Q-balls and Fermi-balls, both of which are typical
nontopological solitons.

We have examined the parameter regions, masses, fluxes, charges, and
energy scales of Q-balls and Fermi-balls that are to be
excluded. These exclusions were derived by analyzing the various
existing and future searches for monopoles, nuclearites, and cosmic
rays. We also analysed the existing results and analyses of the Q-ball
searches of {\it Gyrlyanda} and {\it MACRO}. The experiments
considered here include: ~ (1) underground searches with {\it
Gyrlyanda}, {\it BAKSAN}, {\it Kamiokande}, {\it Super-Kamiokande},
{\it MACRO}, {\it OYA}, {\it MICA}, and {\it AMANDA}; ~(2) searches on
the earth's surface with {\it NORIKURA}, {\it KITAMI}, {\it KEK}, {\it
AKENO}, {\it UCSD}$I\!I$, and {\it TA}; ~ (3) space experiments with
{\it SKYLAB}, {\it AMS}, and {\it OA}.

We summarized these experimental data and obtained bounds on the mass
and the flux of Q-balls and Fermi-balls. Of course, the precise
estimation of bounds should be more carefully made by those who did,
or will do, the experiments.  We believe, however, that our
estimations will give useful information in the research of Q-balls
and Fermi-balls.

\begin{figure}
\begin{center}
\leavevmode
\psfig{file=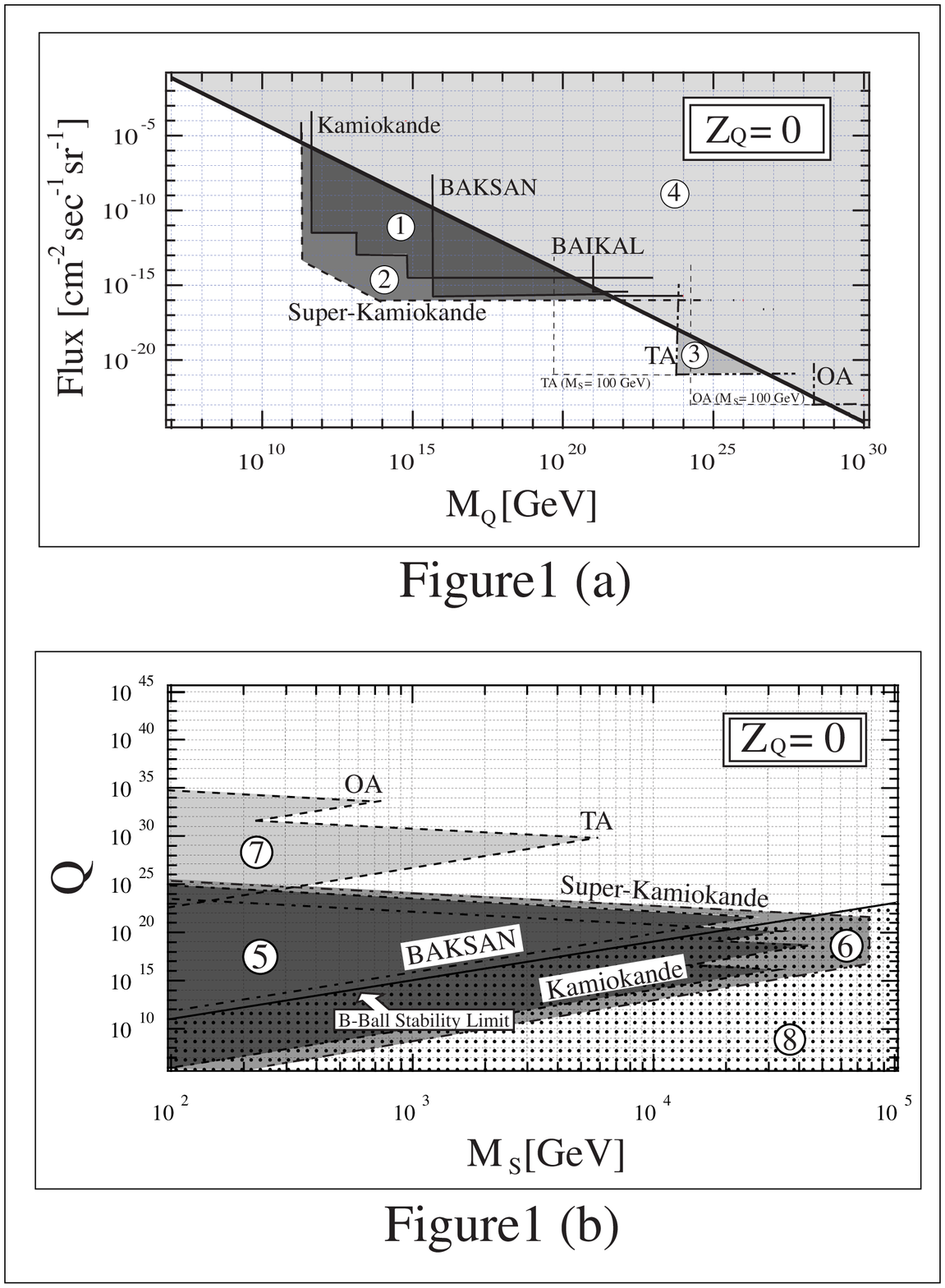,height=17cm}
\vspace*{1cm}
\caption
{Figure\ref{z0com}(a) shows the bounds on flux and mass for neutral
Q-balls ($Z_Q=0$ ), i.e. SENS. The diagonal line shows the flux
expected in case that the Galaxy dark matter ($\sim 0.3 {\rm GeV}
/\mbox{cm}^3$) consists mainly of SENS, and thus the region above this
DM-line ({\bf region 4}), should be excluded. The {\bf region 1} is
excluded by present and past experiments such as {\it
Gyrlyanda}\protect\cite{Bai}, {\it BAKSAN}\protect\cite{Bak}, and {\it
Kamiokande}\protect\cite{Kam}, which give us an accessible analysis
for monopole search, as well as the {\bf region 2} by the {\it
Super-Kamiokande} experiments, the data of which has not been analysed
for the purpose of monopole search. The regions to be searched by
future experiments such as {\it TA}\protect\cite{Ta,Sasaki} and {\it
OA}\protect\cite{Owl}, are also shown ({\bf region 3}).  The lower
mass bound in each experiment is calculated taking $M_S=1\mbox{TeV}$
(taking also $M_S=100$~GeV~ for {\it TA} and {\it OA}).
Figure\ref{z0com}(b) shows bounds on the U(1) charge $Q$ versus the
symmetry breaking scale $M_S$ for SENS ($Z_Q=0$) in the case of the
Galaxy dark matter consisting mainly of SENS (the flux and mass of
SENS in this case is restricted to lie on the diagonal line of
Figure\ref{z0com}(a)).  The {\bf region 5, 6}, and {\bf 7} are
searched in the same experiments as the {\bf region 1, 2}, and {\bf
3}, respectively.  The line with "B-Ball Stability Limit" shows that
the region below it ({\bf region 8}) is not allowed to have the
stability of SENS in case they are "B-balls" (as discussed in the
text). The allowed region $Q > 10^{22}$ is only the upper blank part
of the figure.}
\label{z0com}
\end{center}
\end{figure}

We first investigated Q-balls with an electric charge $Z_Q=0$ (SENS)
which can be detected through proton-decay like process as proposed in
Ref.\cite{Kuz}.  We found that a considerably large region ~$Q ~
\gtrsim~10^{25}$ has already been excluded in $Q-M_S$ plane for $M_S <
100~{\rm GeV}$ only by existing experiments (see
Figure\ref{z0com}(b)), and that a wider region $Q~\gtrsim~10^{35}$
could be searched in future experiments of {\it TA} and {\it OA}. We
also found that the region $Q =B~\gtrsim~10^{22}$ has been excluded
for B-balls for any value of $M_S$.

\begin{figure}
\begin{center}
\leavevmode
\psfig{file=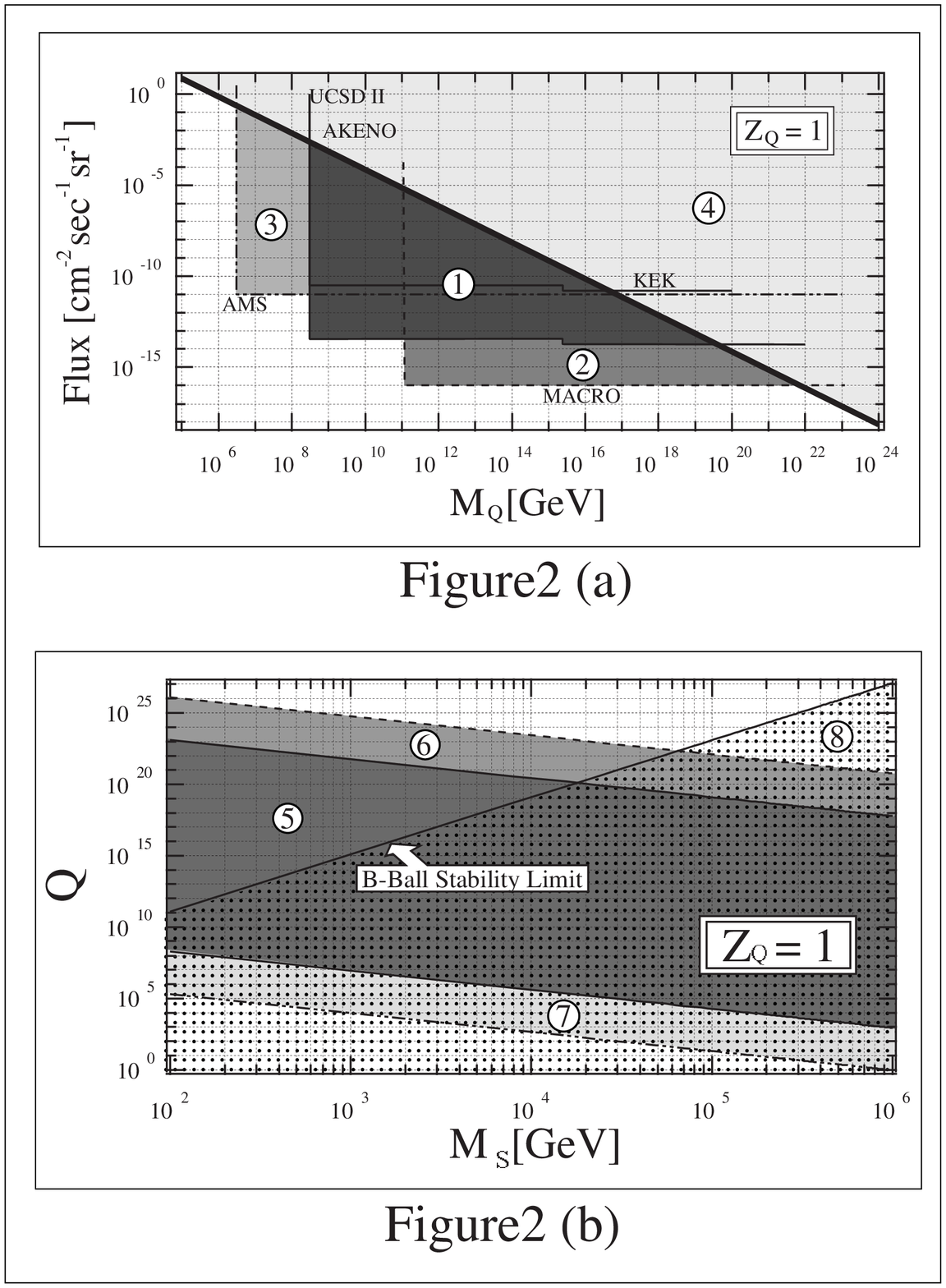,height=19cm}
\vspace*{1cm}
\caption
{Figure\ref{z1com}(a) shows limits on the Q-ball's flux and its mass
for SECS with $Z_Q=1$.  The diagonal line is the same as in
Figure\ref{z0com}(a) and the region above this line ({\bf region 4}),
should be excluded.  The {\bf region 1} is excluded by present and
past experiments such as {\it KEK}\protect\cite{Kek}, {\it
AKENO}\protect\cite{Ake}, and {\it UCSD}$I\!I$\protect\cite{Ucs},
which give us an accessible analysis for monopole search, as well as
the {\bf region 2} by the {\it MACRO}\protect\cite{Gia,Macr,GiaN}
experiments, the data of which has not been analysed for monopole
search yet.  The regions to be searched by future experiments such as
{\it AMS}\protect\cite{Ams}, are also shown ({\bf region 3}).
Figure\ref{z1com}(b) represents bounds on $Q$ and $M_S$ of SECS with
$Z_Q=1$ (see texts for details). The marks and patterns to separate
regions are the same as those used in Figure\ref{z0com}(b).}
\label{z1com}
\end{center}
\end{figure}

\begin{figure}
\begin{center}
\leavevmode
\psfig{file=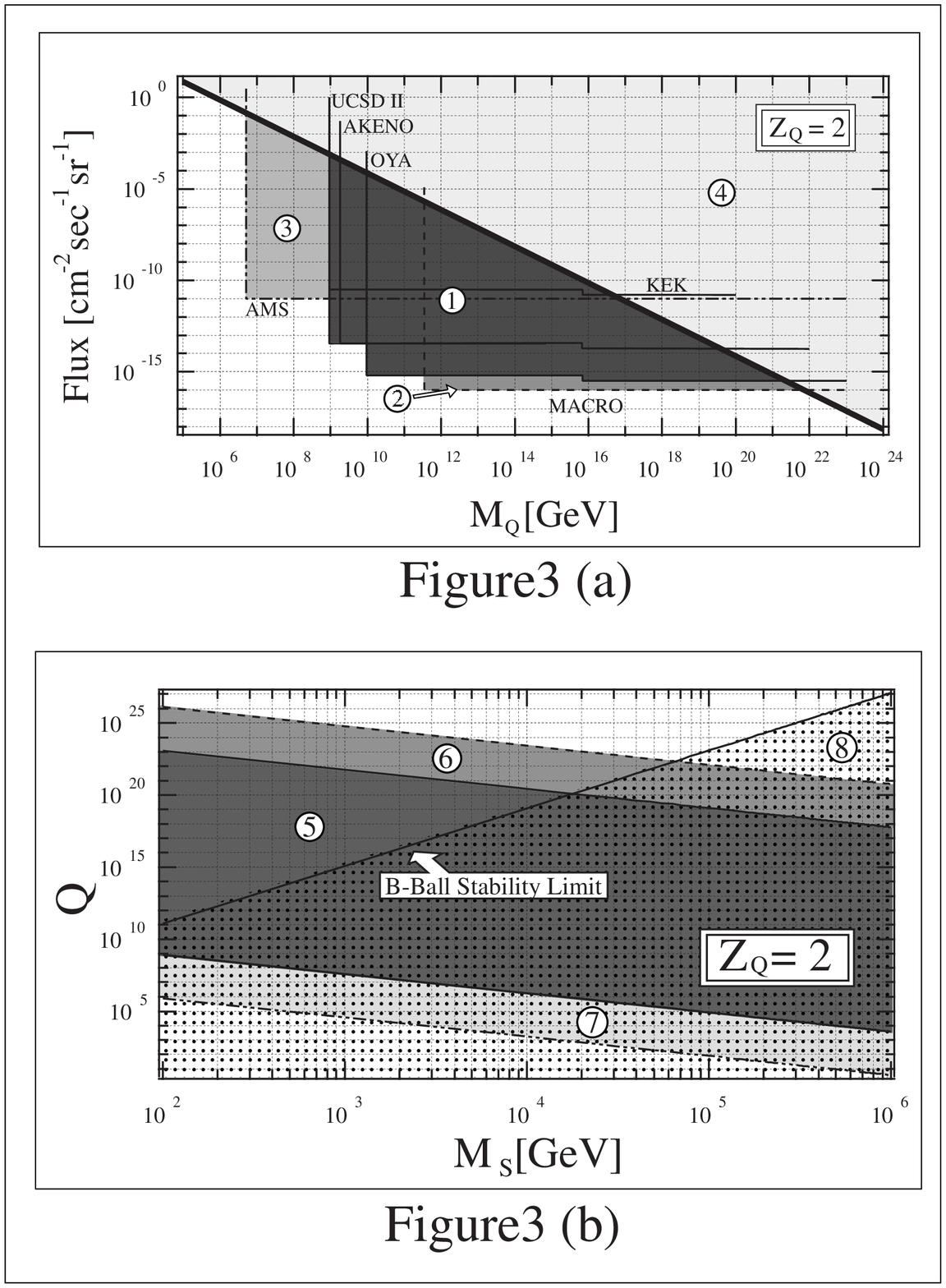,height=21cm}
\vspace*{1cm}
\caption
{Figure\ref{z2com}(a) shows limits on SECS with $Z_Q=2$. This figure
is similar to Figure\ref{z1com}(a) except that the values of the mass
has lower limits and the {\it OYA}\protect\cite{Oya} experiments are
included.  Figure\ref{z2com}(b) represents limits on $Q~ {\rm and} ~
M_S$ of SECS with $Z_Q=2$. The marks and patterns to separate regions
are the same as those used in Figure\ref{z1com}(b).}
\label{z2com}
\end{center}
\end{figure}

\begin{figure}
\begin{center}
\leavevmode
\psfig{file=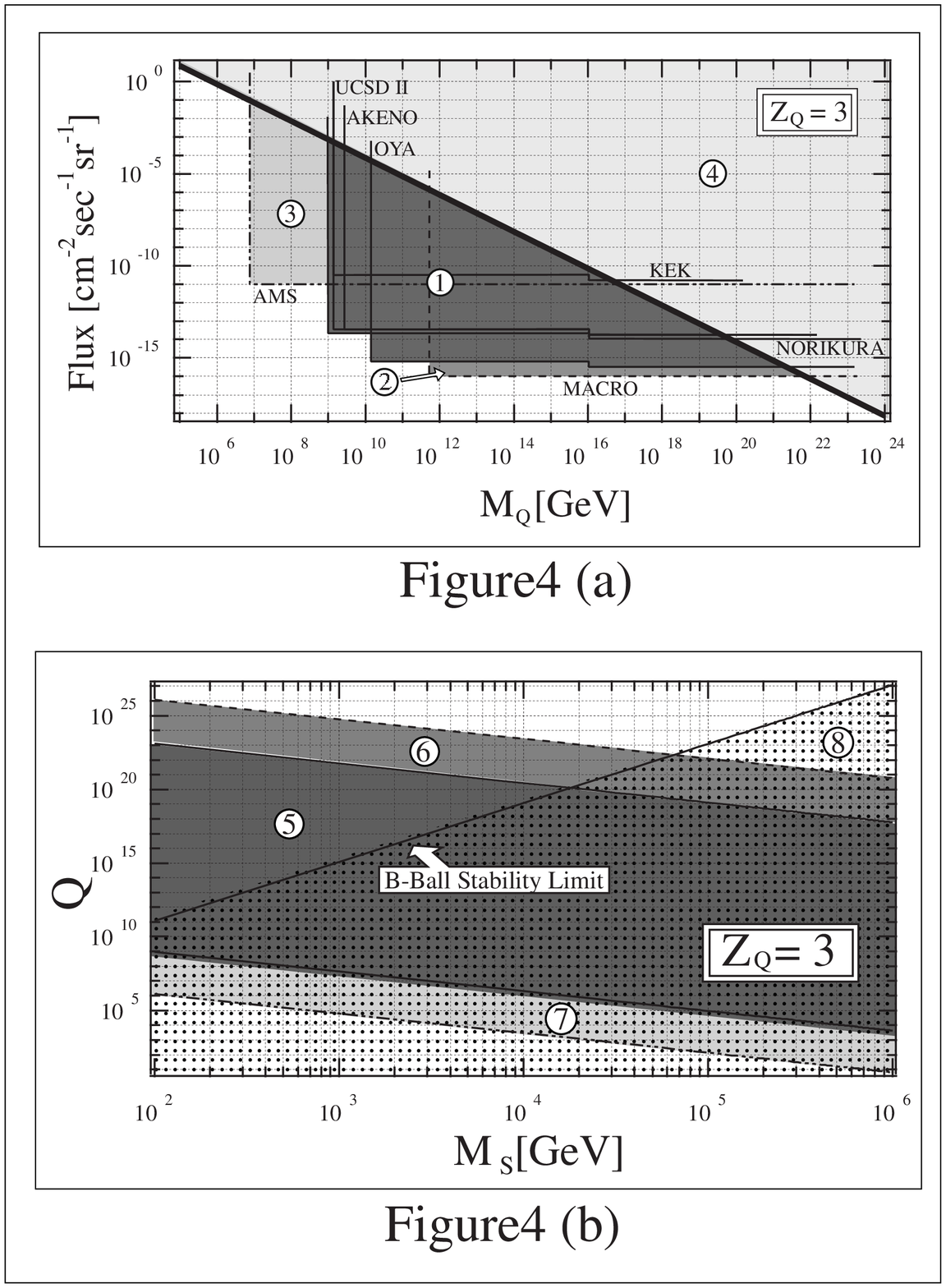,height=21cm}
\vspace*{1cm}
\caption
{Figure\ref{z3com}(a) shows limits on SECS with $Z_Q=3$. This figure
is similar to Figure\ref{z2com}(a) except that the values of the mass
has lower limits and the {\it NORIKURA}\protect\cite{Nor} experiments
are included.  Figure\ref{z3com}(b) represents bounds on $Q~ {\rm and}
~ M_S$ of SECS with $Z_Q=3$.  The marks and patterns to separate
regions are the same as for those used in Figure\ref{z2com}(b).}
\label{z3com}
\end{center}
\end{figure}

\begin{figure}
\begin{center}
\leavevmode
\psfig{file=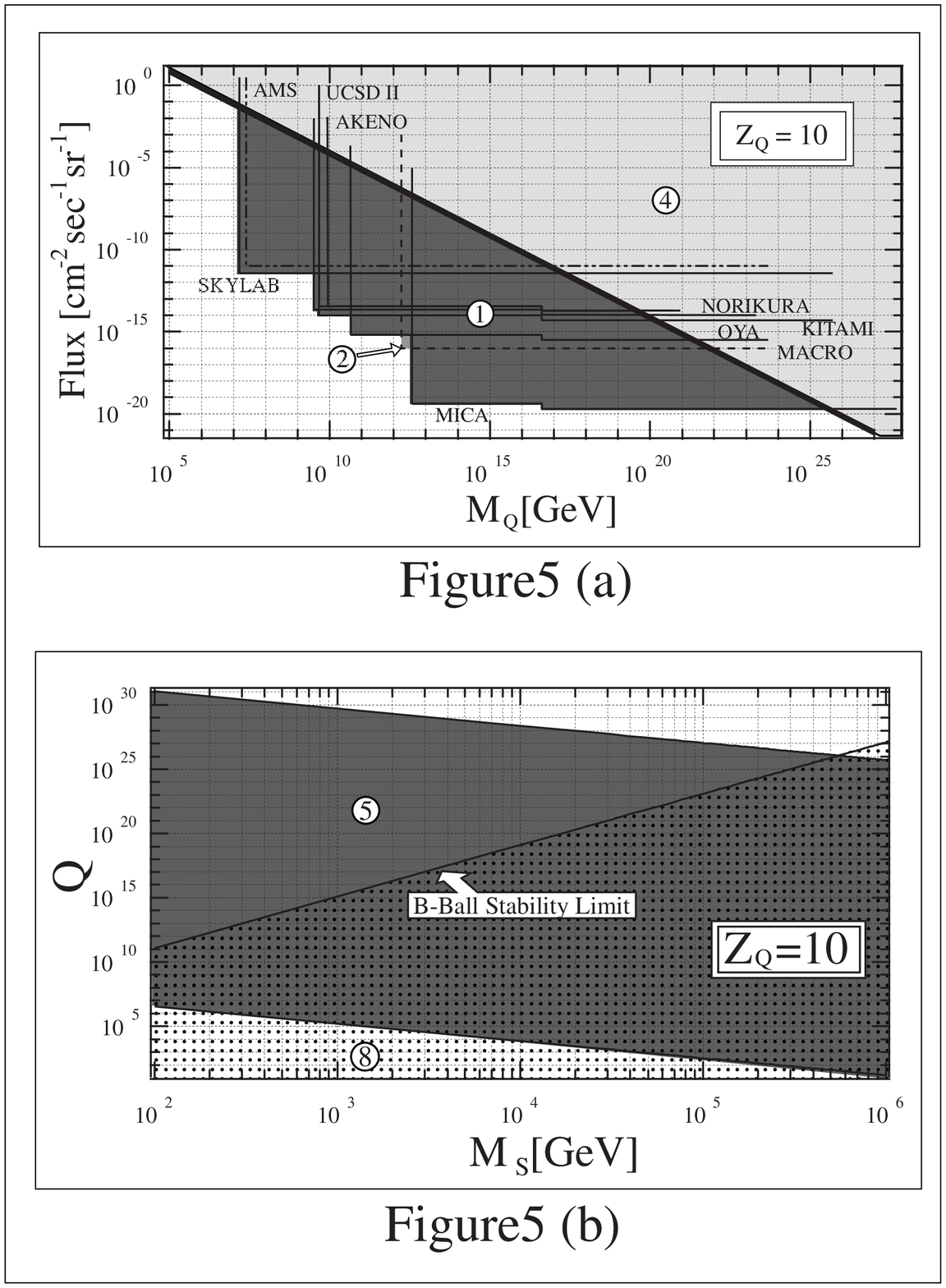,height=20.5cm}
\vspace*{1cm}
\caption
{Figure\ref{z10com}(a) shows limits on SECS with $Z_Q=10$. This figure
is similar to Figure\ref{z3com}(a) except that the values of mass has
lower limits and the experiments of {\it SKYLAB}\protect\cite{Sky},
{\it KITAMI}\protect\cite{Kit}, and {\it MICA}\protect\cite{Mic} are
included.  The experiments of {\it MICA} and {\it SKYLAB} give
stringent limits to the SECS.  Figure\ref{z10com}(b) represents limits
on $Q~ {\rm and}~ M_S$ of SECS with $Z_Q=10$.  The marks and patterns
to separate regions are the same as for those used in
Figure\ref{z3com}(b).}
\label{z10com}
\end{center}
\end{figure}

\begin{figure}
\begin{center}
\leavevmode
\psfig{file=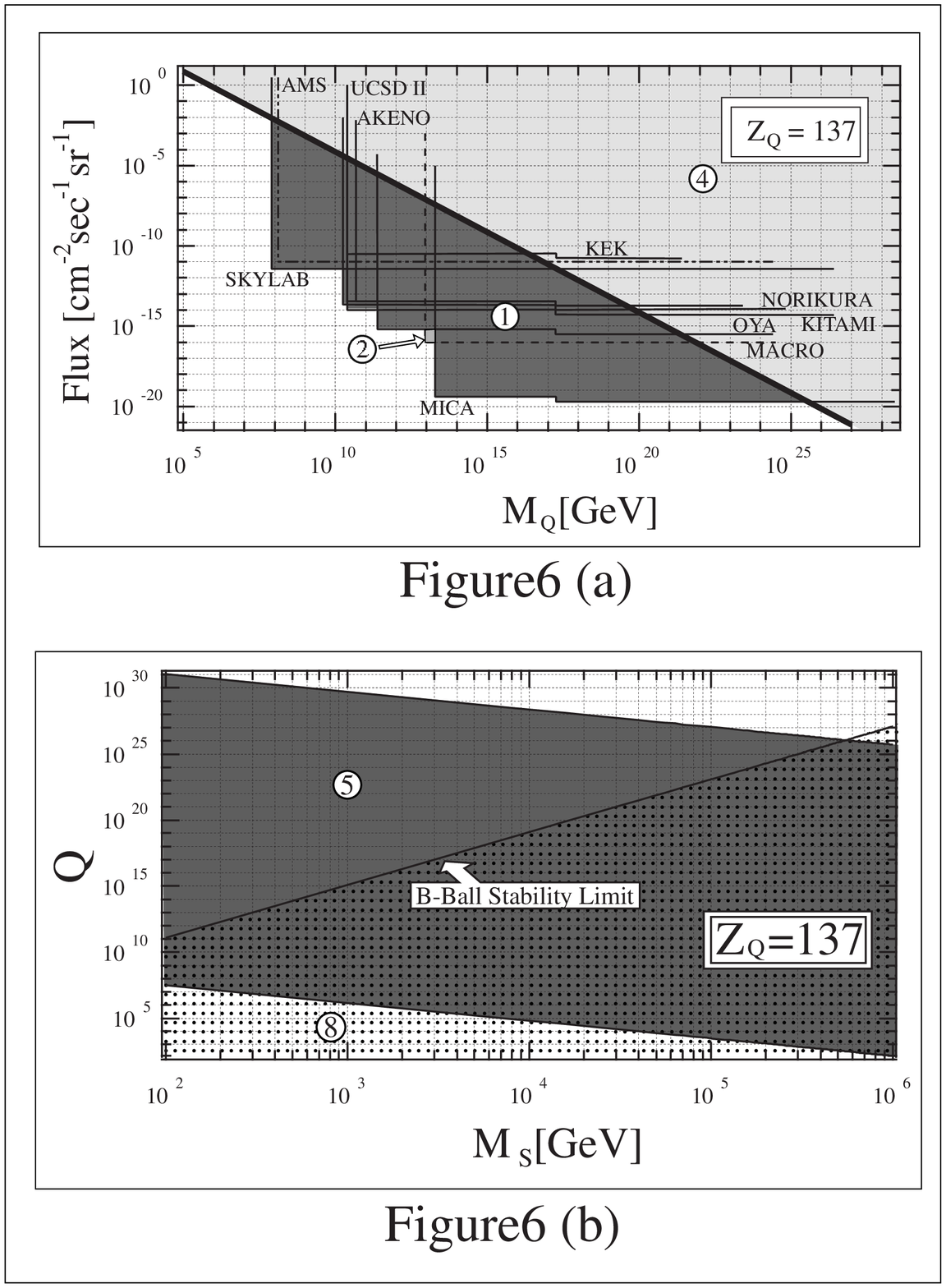,height=20cm}
\vspace*{1cm}
\caption
{Figure\ref{z137com}(a) shows bounds on SECS with $Z_Q=137$. This
figure is similar to Figure\ref{z10com}(a) except for the values of
the excluded mass region.  In this case the cross section for the
collision with matter atoms are assumed to be $\pi R_{eff}^2$ with
$R_{eff} \sim 1 \mbox{\AA}$. We expect that the case where $Z_Q > 137$
is the same as the case where $Z_Q = 137$.  Figure\ref{z137com}(b)
represents limits on $Q~{\rm and} ~M_S$ of SECS with $Z_Q=137$.  The
marks and patterns to separate regions are the same as for those used
in Figure\ref{z10com}(b).}
\label{z137com}
\end{center}
\end{figure}

We next investigated Q-balls with $Z_Q=1,2,3,10, {\rm and} ~137$
(SECS) which interact with matter, mainly by electromagnetic force,
just like nuclearites (though the relations between the radius and the
mass are quite different).  We saw that larger $Z_Q$ gives more
stringent bounds on flux and mass of SECS. We found that for the value
of $M_S \sim 10^2 {\rm GeV}$, experimental data gives more stringent
limitations on SECS global U(1) charge than the stability condition of
B-balls. Q-balls with $Q~\gtrsim~10^{22-26}$ still remain to be
examined (see Figure\ref{z1com}-\ref{z137com}).

\begin{figure}
\begin{center}
\leavevmode
\psfig{file=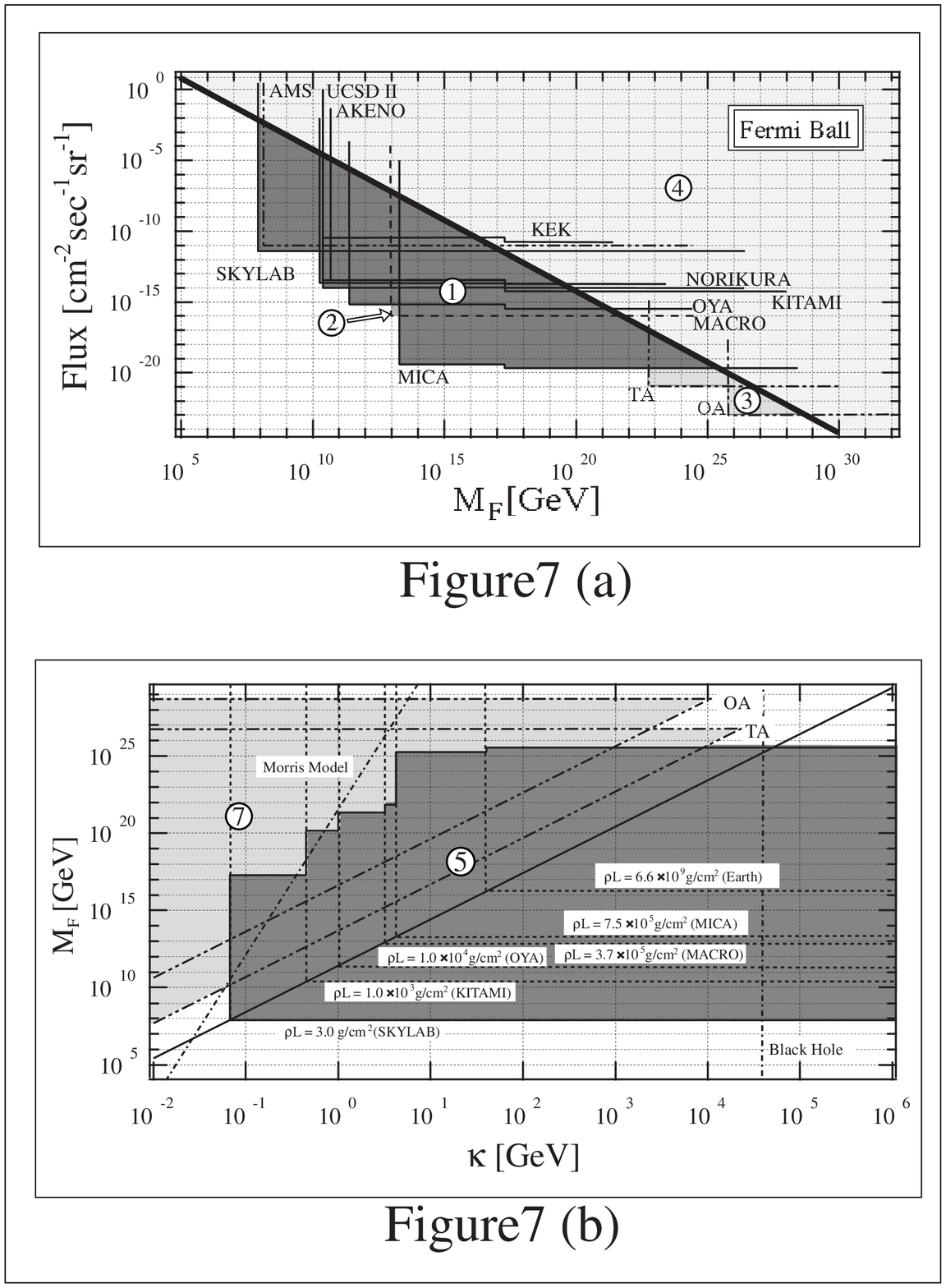,height=20.5cm}
\vspace*{1cm}
\caption
{Figure\ref{fermicom}(a) shows the limits of flux and mass of
Fermi-balls obtained from various experiments. This is similar to
Figure\ref{z137com}(a) (SECS with $Z_Q =137$), although it has
additional restrictions to be expected in the future {\it TA} and {\it
OA} experiments.  Figure\ref{fermicom}(b) represents an excluded
region in the $M_F-\kappa$ plane. The {\bf region 5} is excluded by
the available data of present and past experiments. The {\bf region 7}
is to be investigated by future experiments such as {\it TA} and {\it
OA}. The Morris's Model line(Eq.(\ref{mmodel})) and black hole
limit(Eq.(\ref{bh2})) are also shown in this figure (taking $C=1$). }
\label{fermicom}
\end{center}
\end{figure}

We finally investigated Fermi-balls with electric charge $Z_F \gg
137$, which are expected to be rather stable against perturbative
deformation and fragmentation.  We obtained bounds on mass of
Fermi-balls, $M_F~\lesssim~10^{8}$ GeV ~and $M_F~\gtrsim~10^{29}$ GeV
for $\kappa ~\gtrsim ~0.1$ GeV in $M_F-\kappa$~plane. If we further
assume the Morris Model\cite{Mor}, we obtained more stringent
constraints.  We also noted the importance of future {\it TA} and {\it
OA} experiments in the search for Fermi-balls as seen in
Figure\ref{fermicom}(b).

We lastly note that the charged solitons with a relatively light mass,
$M~\lesssim~ 10^8$GeV, on which we have not focussed our attention,
have astrophysical difficulties.  If we consider that the dark matter
consists mainly of such solitons with large cross section with matter
and radiation, we will be faced by difficulties including `the
Galaxy-halo in-fall', `too much heating of disk molecules', and `too
small density fluctuations in the early universe'\cite{Sta,Der}.
Therefore, We must show an interest in the window for the heavier
charged solitons.

From these experimental bounds, we comprehensively obtained the
stringent limitations on the properties of Q-balls and the Fermi-balls
and then noted the possible importance of future experiments of {\it
TA} and {\it OA}.  These bounds will help us study the unsolved
cosmological problems that we have mentioned in this paper, by
developing more realistic scenarios. In these cases, Q-balls and/or
Fermi-balls play an essential role.\\[1cm]

{\bf Acknowledgement}:
\\
We thank H.Tawara, M.Fukugita, M.Fukushima, and M.Sasaki for useful
comments.


\newpage
%
%

%
%
%
%
%
%
%
%
%
%

%
%

\end{document}